\documentclass[journal]{IEEEtran}

\normalsize

\usepackage{amsmath,amssymb,graphicx,fancyvrb,color}

\DeclareMathOperator*{\argmin}{\arg\!\min}

\begin{document}
%
\title{Iterative pre-distortion of the non-linear satellite channel}
%
%
%

\author{Thibault~Deleu,
        Mathieu~Dervin,
        Kenta~Kasai,
        and~Fran\c{c}ois~Horlin
\thanks{T. Deleu* and F. Horlin are with the OPERA - Wireless Communications Group, Universit\'e Libre de Bruxelles, Brussels, Belgium.}
\thanks{M. Dervin is with Thales Alenia Space, Toulouse, France.}
\thanks{K. Kasai is with Tokyo Institute of Technology, Tokyo, Japan.}}

\markboth{IEEE Transactions on Communications}%
{Submitted paper}
\maketitle
\begin{abstract}
Digital Video Broadcasting - Satellite - Second Generation (DVB-S2) is the current European standard for satellite broadcast and broadband communications. 
It relies on high order modulations up to $32$-amplitude/phase-shift-keying (APSK) in order to increase the system spectral efficiency. 
Unfortunately, as the modulation order increases, the receiver becomes more sensitive to physical layer impairments, and notably to the distortions induced by the power amplifier and the channelizing filters aboard the satellite. 
Pre-distortion of the non-linear satellite channel has been studied for many years. 
However, the performance of existing pre-distortion algorithms generally becomes poor when high-order modulations are used on a non-linear channel with a long memory. 
In this paper, we investigate a new iterative method that pre-distorts blocks of transmitted symbols so as to minimize the Euclidian distance between the transmitted and received symbols. 
We also propose approximations to relax the pre-distorter complexity while keeping its performance acceptable. 
\end{abstract}

\begin{IEEEkeywords}
Pre-distortion, non-linear satellite channel, DVB-S2.
\end{IEEEkeywords}
\IEEEpeerreviewmaketitle
\section{Introduction}\label{intro}
In broadcast or broadband satellite communication, information is most often exchanged between one hub and many user terminals in a so-called star topology. We focus here on the forward link, defined as the link from the hub towards the user terminals, while the return link (when it exists) refers to the link from a user terminal towards the hub. In such context, the available radio spectrum is generally divided into sub-bands, also referred to as channels, which are separately amplified by different power amplifiers aboard the satellite. In a single carrier per channel scenario, each carrier transmitted on the forward link is separately amplified by a different power amplifier. As a single carrier signal shows limited envelope variations, this conveniently allows each power amplifier to be driven close to its saturation point, so that the power consumption aboard the satellite is minimized. When the link budget is good enough, it is possible to increase the spectral efficiency of the system by using high-order modulations. However, the transmission channel includes non-linear inter-symbol interference (ISI) due to the combination of the non-linear high power amplifier (HPA) aboard the satellite with linear filtering present in the channel. Moreover, the larger the modulated carrier bandwidth, the more interference occurs due to the bandpass nature of the onboard channelizing filters. Higher-order modulations being more  sensitive to the non-linear ISI, compensation algorithms are necessary to remove the non-linear interference induced by the satellite channel, and fully benefit from the spectral efficiency improvement.\\

In the literature, the methods proposed to compensate for the non-linear interference can be divided into two categories: equalization and pre-distortion. 

Firstly, the non-linear interference can be mitigated with an equalizer at the receiver side. If the channel is exactly known, the maximum-a-posteriori (MAP) symbol detection algorithm and the alternative maximum-likelihood sequence detection algorithm can be perfectly defined. 
However, the complexity of these optimum algorithms increases exponentially with the channel length and modulation order, so that several sub-optimum algorithms have been proposed in the literature.
For instance, in~\cite{channelshortening}, the detection of the received signal is based on a reduced channel model described in~\cite{novelsiso} combined with a channel shortening technique described in~\cite{optimal_channel}. Adaptive non-linear equalizers have been proposed in \cite{guti}, \cite{regression}, where a-priori channel knowledge is not required. 
In \cite{olmos}, joint equalization and channel decoding is performed using Gaussian processes. 
To take advantage of the channel coding gain, iterative turbo-equalization structures have also been considered \cite{burnet}, \cite{reduced_turbo}. 

Secondly, the channel non-linear interference can be compensated by pre-distortion at the transmitter side. This approach is particularly interesting in the forward link of a broadband satellite system, where it is preferred to concentrate the computational load in the hub and relax as much as possible the complexity of the terminals. One usually uses the term \textit{signal} (or \textit{waveform}) pre-distortion when it is located \textit{after} the pulse shaping filter. This kind of pre-distortion can be applied to compensate memoryless channels,  as shown in \cite{linear1}, \cite{linear4} and references therein. The pre-distorter is then an approximation of the inverse characteristic of the power amplifier at the transmitter side. This method 
 can be analog or digitally implemented (see for example the adaptive implementations in \cite{nn_tables} and \cite{nn_predistorter}). 
 On the other hand, we refer to \textit{data} pre-distortion when a pre-distortion of the data symbols is applied \textit{prior} to the pulse shaping. This allows compensating for ISI and avoids out-of-band emissions. A first approach is to consider a pre-distorter based on the Volterra model, a common tool to describe the input-output relation of a non-linear system with memory, as described in \cite{schetzen:volterra}. 
 The coefficients of the pre-distorter are adaptively determined to minimize the mean-square error (MSE), as in~\cite{adapt_volterra1},~\cite{adapt_volterra2}. The complexity of such pre-distorters may be high and the convergence of adaptive algorithms may be slow, \color{black}so that pre-distortion methods based on reduced Volterra models have been studied in \cite{polynomials} and references therein. 
The order-$p$ inverse for non-linear systems has been described in \cite{p_inverse} and applied to the satellite channel in \cite{principles}. The order-$p$ pre-distorter removes, up to the order $p$, all Volterra terms from the channel model relating the received to the transmitted symbols (note that this algorithm can actually be applied to both pre-distortion and equalization). 
Another structure of interest relies on a look-up table (LUT). 
In \cite{data_qam}, the value of each pre-distorted symbol is a function of the neighboring initial symbols, which can be calculated offline and stored in a LUT. 
The pre-computation of these values aims at minimizing the MSE between the initial and the received symbols. The performance of this algorithm has been assessed for high-order modulations in \cite{Casini:modem}.\\

Except for the order-$p$ inverse, existing pre-distortion methods suffer from a performance loss due to their intrinsic structure: 
pre-distorters based on a Volterra structure cannot cope with the huge number of coefficients required to theoretically represent the channel inverse and are most of the time limited in order and memory. 
Pre-distorters based on LUT have a number of entries exponentially growing with the modulation order so that the pre-distorter length must be limited. 
In case of large channel length and high HPA non-linearity, these pre-distorters are therefore expected to perform poorly. 
This is also the case for the order-$p$ pre-distorter since it creates higher order terms, that become more powerful than the cancelled terms.
In this paper, we propose an iterative pre-distortion algorithm, which can be seen as a pre-distorter of infinite order and finite length. Very large pre-distorter lengths can be considered because the algorithm complexity computed per symbol is independent of this parameter. 
Improved performance is therefore expected compared to state-of-the-art pre-distortion methods. \color{black}
The proposed scheme independently pre-distorts successive symbol blocks. 
To pre-distort each block of symbols, an iterative algorithm is used, aiming at minimizing the Euclidian distance between the initial symbol and the received symbol sequence. 
Based on the system model defined in Section~II, we describe the proposed algorithm in Section~III. A main concern of the algorithm design is its complexity, so that variations of the algorithm of much lower complexity are proposed in Section IV. 
The complexity and the performance of the different algorithms are compared in Sections V and VI respectively.

\section{System model}
\subsection{Satellite Channel}
\begin{figure*}
\center
\includegraphics[width=0.9\textwidth]{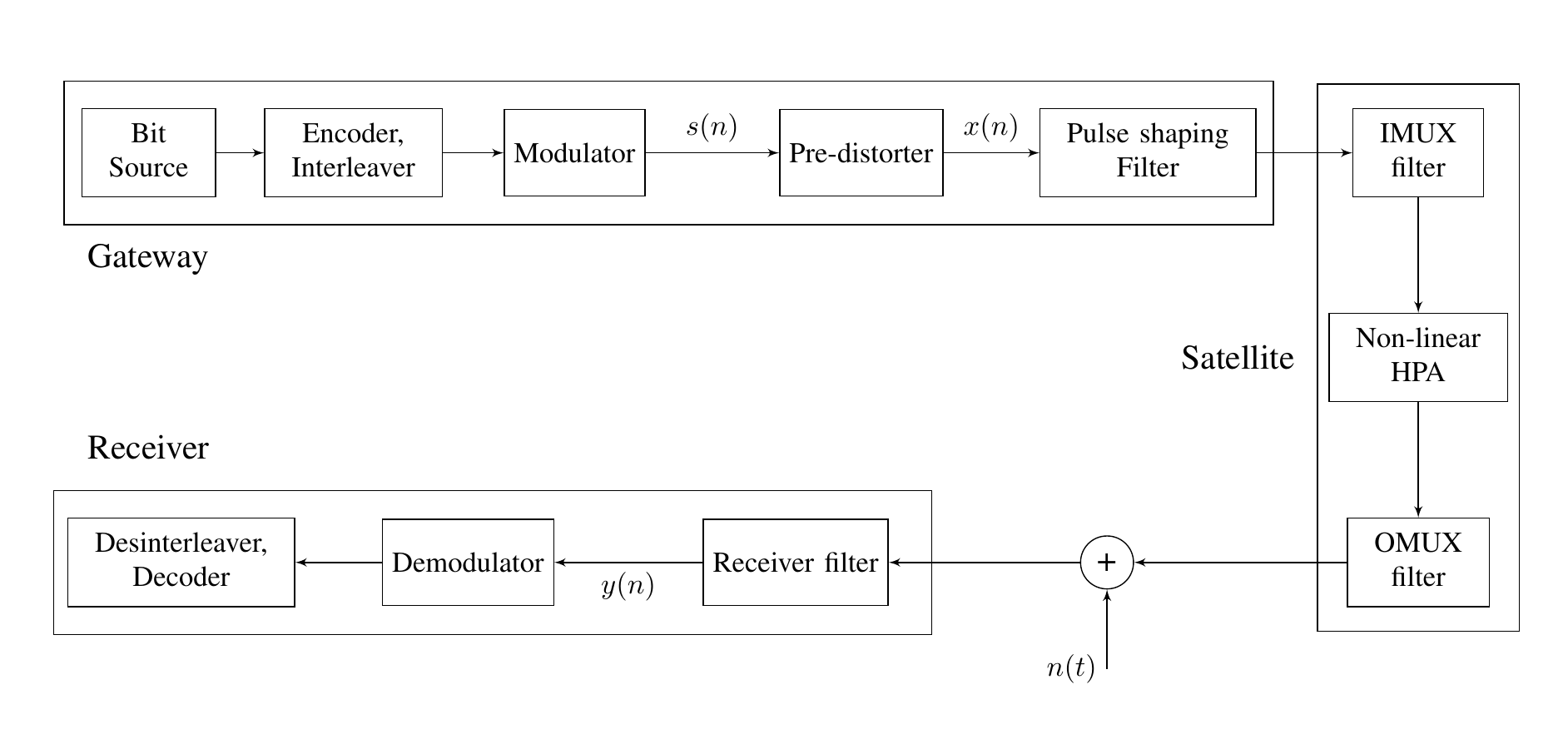}
\caption{\large Block diagram of the satellite channel}
\label{fig1}
\end{figure*}

A block diagram of the satellite channel is depicted in Fig.~\ref{fig1}. 
At the transmitter, data bits are first encoded, interleaved and linearly modulated. 
In this work, we will only consider the highest modulation order defined in the DVB-S2 standard (\cite{standard}): the 32-amplitude/phase-shift-keying (APSK) modulation. 
Based on the data symbols, denoted as $s(n)$, the pre-distorter produces the pre-distorted symbols, denoted as $x(n)$. 
The pre-distorted symbols are shaped with a square-root raised cosine (SRRC) filter and the resulting signal is transmitted to the satellite.
At the satellite, the input multiplexer (IMUX) filter is a bandpass filter that selects the sub-band to be amplified. 
The satellite HPA can be seen as a non-linear memoryless device. 
The output multiplexer (OMUX) filter is also a bandpass filter, necessary to remove the out-of-band components produced by the power amplifier. 
At the receiver, the signal is filtered with a SRRC filter and sampled to produce the received samples $y(n)$. 
The demodulator performs a memoryless detection, assuming that residual interference after pre-distortion behaves like additive white Gaussian noise (AWGN). 
The demodulator produces a-posteriori statistics of the encoded bits, which are transmitted to the decoder after desinterleaving. 
The pre-distortion block is assumed to have a perfect knowledge of the channel, and is dedicated to the mitigation of the non-linear ISI induced by the combination of the linear filters and the HPA. 
The pre-distortion block is further detailed in the next section. 
\color{black}

\subsection{Volterra Model}
The Volterra model is an analytical model that describes the relation between the input and the output of a non-linear system with memory. The case of the baseband non-linear satellite channel has been described in \cite{principles}. The relation between the pre-distorted symbols $x(n)$ at the channel input and the received symbols $y(n)$ is given by:
\begin{align} \label{eq:Volt}
y(n)=& \sum_{m=0}^{\infty}\sum_{n_1...n_{2m+1}}H_{2m+1}(n_1...n_{2m+1})x(n-n_1)...\nonumber\\&x(n-n_{m+1})x^*(n-n_{m+2})...x^*(n-n_{2m+1})+w(n). 
\end{align}
The coefficients $H_{2m+1}(n_1...n_{2m+1})$ are called the Volterra kernels of the system. 
The first sum in~(\ref{eq:Volt}) represents the different orders of the non-linearity induced by the power amplifier. The second set of sums represents the memory of the system, which is theoretically infinite. In practice however, the length of the channel can be reasonably assumed of finite length. We denote the anti-causal memory of the channel as $L_1$ and the causal memory of the channel as $L_2$. The total channel length is then denoted as $L_\text{c}=L_1+L_2+1$. In~(\ref{eq:Volt}), each index $n_i$  varies thus from $-L_1$ to $L_2$. The received symbols are also corrupted by thermal noise $w(n)$, which is supposed to behave like AWGN.\color{black}

\subsection{Total degradation}\label{sec:td}
The performance of pre-distortion or equalization algorithms in a non-linear satellite channel is usually quantified in terms of the total degradation~\cite{data_qam}, \cite{Casini:modem}, \cite{beidas:2carriers}. The total degradation, denoted as TD, is defined as follows:
\begin{align}\label{eq:TD}
\text{TD}[dB]=&\text{OBO[dB]}+L^\text{omux}\text{[dB]}\nonumber\\&+\left[\frac{Eb}{N_0}\right]^\text{NL}_\text{req}\text{[dB]}-\left[\frac{Eb}{N_0}\right]^\text{AWGN}_\text{req}\text{[dB]},
\end{align}
where OBO is the HPA power backoff, $L^\text{omux}$ is the mean power loss in the OMUX filter, $\left[\frac{Eb}{N_0}\right]^\text{NL}_\text{req}$ and $\left[\frac{Eb}{N_0}\right]^\text{AWGN}_\text{req}$ are the average symbol energy over noise ratio required to achieve a given bit error rate (BER) or frame error rate (FER), in the non-linear and AWGN channels. 
As shown by~(\ref{eq:TD}), the total degradation depends on the OBO, and the optimum OBO which minimizes the total degradation can significantly be different depending on the considered pre-distortion technique, as shown in~\cite{data_qam}, \cite{Casini:modem}, \cite{beidas:2carriers}. Pre-distortion techniques must therefore be compared based on the minimum total degradation they can reach.
\color{black}
\section{Per-block iterative pre-distortion}\label{Sec3}

\subsection{Minimization of Euclidian Distance}\label{sec:MED}
We consider the pre-distortion of length-$N$ symbol blocks, assuming that the transmitter has perfect knowledge of the channel. \color{black} 
For a given block, we denote by $\bold{s}$ the vector with elements comprising a symbol block: $\bold{s}=[s(1)...s(N)]$. For each symbol block, the pre-distorter produces a modified symbol block of length $N$, denoted by the vector $\bold{x}=[x(1)...x(N)]$. At the receiver, $N$ samples are also gathered in a vector of size $N$, denoted by $\bold{y}=[y(1)...y(N)]$. In addition, we denote by $\bold{y(x)}$ the vector $\bold{y}$ of the received symbols when the block $\bold{x}$ is sent at the channel input. In this section, we propose an algorithm that precodes the block $\bold{x}$ so that $||\bold{y}-\bold{s}||_2$ is minimized. 
 
Since the optimal compensation of a finite length channel is of infinite length, it is important to take $N$ as large as possible. 
However, $N$ cannot be too large to prevent too high latency. 
In this work, we take $N$ equal to the number of symbols in the physical layer frame as defined in the DVB-S2 standard (a few thousand symbols). 
\color{black}
Note that the length of the block $\bold{y}$ should be equal to $N+L_\text{c}-1$. 
However, memoryless detection is  applied on consecutive received symbols, so that the symbols $y(-L_1),\hspace{1mm}...\hspace{1mm},\hspace{1mm} y(-1)$ and $y(N+1),\hspace{1mm}...\hspace{1mm},\hspace{1mm}y(N+L_2)$ can be neglected. The pre-distorter minimizes the Euclidian distance assuming a noiseless channel. Since the AWGN is independent of the transmitted sequence, this also minimizes the MSE at the receiver in presence of AWGN. \color{black}The vector $\bold{y}$ can be developed using~(\ref{eq:Volt}). However, there is no straightforward derivation of the block $\bold{x}$ that minimizes $||\bold{y}-\bold{s}||_2$. We therefore propose an iterative algorithm to determine the pre-distorted block $\bold{x}$. Each iteration of the algorithm is divided into $N$ steps, respectively focused on consecutive symbols of the block of interest. The pre-distorted block after Step $j$ of Iteration $k$ is denoted as $\bold{x_\textit{k,j}}=[x_{k,j}(1)... x_{k,j}(N)]$, where only the $j\text{th}$ pre-distorted value is modified and is chosen to minimize $||\bold{y}-\bold{s}||_2$ when $\bold{x_\textit{k,j}}$ is transmitted. All other pre-distorted values are thus kept equal to their values from the previous step.  For each iteration, $x_{k,j}(n)$ is mathematically expressed as follows. For the first step ($j=1$),
\begin{align}\label{eq:Def}
x_{k,1}(n)=\begin{cases}&x_{k-1,N}(n),\hspace{5mm}n\neq 1,\\&\underset{x_{k,1}(1)}{\argmin}[||\bold{y}-\bold{s}||_2\big{|}\forall i\neq 1: x(i)=x_{k-1,N}(i)],\\&\hspace{55mm}n=1,\end{cases}
\end{align}
and thereafter ($j>1$),
\begin{align}\label{eq:spec}
x_{k,j}(n)=\begin{cases}&x_{k,j-1}(n),\hspace{5mm}n\neq j,\\&\underset{x_{k,j}(j)}{\argmin}[||\bold{y}-\bold{s}||_2\big{|}\forall i\neq 1: x(i)=x_{k,j-1}(i)],\\&\hspace{35mm}n=j.\end{cases}
\end{align}
Note that $x_{k,1}(n)$ is calculated using the end values of the previous iteration, except for the first iteration where $x_{k,1}(n)$ is calculated using the un-pre-distorted values. The vector $\bold{\epsilon_\text{k,j}}$ is defined as the difference between $\bold{y}$ and $\bold{s}$ when the sequence obtained after Step $j$ of Iteration $k$ is transmitted:
\begin{align}
\bold{\epsilon_\text{k,j}}\triangleq \bold{y}-\bold{s}\big{|}\forall i: x(i)=x_{k,j}(i).
\end{align}
By definition of the algorithm, we have:
\begin{align}\label{eq:dec}
||\bold{\epsilon_\text{k,j}}||_2\leq||\bold{\epsilon_\text{k,j-1}}||_2,
\end{align}
so that the convergence of proposed algorithm is ensured. 
The term $||\bold{y}-\bold{s}||_2$ minimized in~(\ref{eq:Def}) can be seen as a non-linear function of the complex variable $x_{k,j}(n)$. The coefficients of this function can be found using the Volterra model and depend on the fixed pre-distorted values in~(\ref{eq:Def}). Since the channel has finite length,~(\ref{eq:Def}) can be simplified as:
\begin{align}\label{eq:Erreur}
x_{k,j}(j)=&\underset{x_{k,j}(j)}{\argmin}[||\bold{y}-\bold{s}||_2\big{|}\forall i\neq j: x(i)=x_{k,j-1}(i)]\nonumber\\
= &\hspace{2mm}\underset{x_{k,j}(j)}{\argmin}[\sum_{m=max(1, \hspace{1mm} j-L1)}^{min(N, \hspace{1mm}j+L2)}{|y(m)-s(m)|^2}\big{|}\forall i\neq j: \nonumber\\&\hspace{40mm}x(i)=x_{k,j-1}(i)].
\end{align}
The complexity of the algorithm is very high since it is necessary to successively find the minimum of $N$ complex non-linear functions for each iteration. 
Moreover, the number of Volterra coefficients in each equation can be very high in the case of high-order non-linearities. Therefore, the pre-distorted symbols defined in~(\ref{eq:Def}) are difficult to compute in practice. 
In the next subsection, we propose an algorithm of much lower complexity to compute the pre-distorted symbols. 
We refer to this algorithm as the \textit{small-variation} algorithm. 

\subsection{Small-Variation Algorithm}
The small-variation algorithm has the same iterative structure as the algorithm presented in the previous subsection. However, at Step $j$ of Iteration $k$, it calculates a suboptimal value for $x_{k,j}(j)$ in a much less complex way.
We first define ${\Delta}_{k,j}$ as:
\begin{align} \label{eq:Delta}
x_{k,j}(j)=x_{k,j-1}(j)+\Delta_{k,j}
\end{align}
Thus, the variation from $x_{k,j-1}(j)$ to $x_{k,j}(j)$ is considered as the unknown variable instead of $x_{k,j}(j)$ itself. The case $j=1$ is not explicitly given anymore in the following derivations, as it is always similar to~(\ref{eq:Def}). The vector $\bold{\Delta_\textit{k,j}}$ is defined as a zero vector of length $N$, except for the element $j$ which is equal to $\Delta_{k,j}$, so that:
\begin{align}
\bold{x_\textit{k,j}}=\bold{x_\textit{k,j-1}}+\bold{\Delta_\textit{k,j}}.
\end{align} 
We define the value $\Delta^\text{opt}_{k,j}$ as the optimum value that minimizes~(\ref{eq:spec}):
\begin{align}\label{eq:deltaopt1}
\Delta^\text{opt}_{k,j}=\underset{\Delta_{k,j}}{\argmin}[||\bold{y}-\bold{s}||_2&\big{|}\forall i\neq j: x(i)=x_{k,j-1}(i),\nonumber\\&x_{k,j}(j)=x_{k,j-1}(j)+\Delta_{k,j}]
\end{align}
It is possible to simplify~(\ref{eq:deltaopt1}) as in~(\ref{eq:Erreur}), but we prefer to adopt the following more compact notation:
\begin{align}\label{eq:deltaopt}
\Delta^\text{opt}_{k,j}&=\underset{\Delta_{k,j}}{\argmin}[||\bold{y}(\bold{x_\textit{k,j-1}}+\bold{\Delta_\textit{k,j}})-\bold{s}||_2].
\end{align}
We define:
\begin{align} \label{eq:Out}
\bold{F_\textit{k,j}^\text{NL}}\triangleq \bold{y}(\bold{x_\textit{k,j-1}}+\bold{\Delta_\textit{k,j}})-\bold{y}(\bold{x_\textit{k,j-1}})
\end{align}
so that:
\begin{align}\label{eq:perfect}
\Delta^\text{opt}_{k,j}&=\underset{\Delta_\textit{k,j}}{\argmin}[||\bold{y}(\bold{x_\textit{k,j-1}})-\bold{s}+\bold{F_\textit{k,j}^\text{NL}}||_2]\nonumber\\
&=\underset{\Delta_\textit{k,j}}{\argmin}[||\bold{\epsilon_\textit{k,j-1}}+\bold{F_\textit{k,j}^\text{NL}}||_2].
\end{align}
Each element ${F}_{k,j}^\text{NL}(n)$ represents the output $n$ variation resulting from a variation of the input symbol j at Step $j$ during Iteration $k$. The vector $\bold{F}_{k,j}^\text{NL}$ can be seen as a vector of functions depending on the scalar variable $\Delta_{k,j}$. Inspecting~(\ref{eq:Volt}) and (\ref{eq:Out}), it can be mathematically computed that each element $F_{k,j}^\text{NL}(n)$ takes the form:
\begin{align} \label{eq:FNL}
{F_{k,j}^\text{NL}}(n)=\begin{cases}&0,\hspace{5mm}n<j-L_2, n>j+L_1,\\&\sum_{m_1=0}^\infty{\sum_{m_2=0}^\infty{A^{n}_{k,j}(m_1,m_2)\Delta_{k,j}^{m_1}(\Delta_{k,j}^*)^{m_2}}},\\&\hspace{30mm}n\geq j-L_2, n\leq j+L_1,\end{cases}
\end{align}

where the coefficients $A^{n}_{k,j}(m_1,m_2)$ depend on the Volterra coefficients and the sequence of pre-distorted symbols. For the sake of clarity, Appendix A gives some examples for the coefficients $A^{n}_{k,j}(m_1,m_2)$ in the case of simple Volterra models consisting of only a single Volterra coefficient. In the general case of a channel depending on several Volterra coefficients, the value of $A^{n}_{k,j}(m_1,m_2)$ can be obtained by first computing the value of $A^{n}_{k,j}(m_1,m_2)$ corresponding to each Volterra coefficient taken independently and then summing all the obtained values.\\  
The small-variation algorithm is based on the assumption that each function $F_{k,j}^{NL}(n)$ can be approximated by keeping only its linear dependency on $\Delta_{k,j}$: 
\begin{align} \label{eq:approx sv}
F_{k,j}^\text{NL}(n)\approx F_{k,j}^\text{Lin}(n)\triangleq A^{n}_{k,j}(1,0)\Delta_{k,j}+A^{n}_{k,j}(0,1)\Delta_{k,j}^*.
\end{align} 
This will be more likely the case for small values of $\Delta_{k,j}$. Denoting $\bold{F}_{k,j}^\text{Lin}(\Delta_{k,j})$, $\bold{A}_{k,j}(1,0)$, and $\bold{A}_{k,j}(0,1)$ the vectors obtained with elements $F_{k,j}^\text{Lin}(n)$, $A^{n}_{k,j}(1,0)$ and $A^{n}_{k,j}(0,1)$, with $n$ varying from $1$ to $N$, we have:
\begin{align}\label{eq:yoyo}
\bold{F}_{k,j}^\text{NL}\approx \bold{F}_{k,j}^\text{Lin}\triangleq \bold{A}_{k,j}(1,0)\Delta_{k,j}+\bold{A}_{k,j}(0,1)\Delta_{k,j}^*
\end{align}
Instead of calculating the value $\Delta^\text{opt}_{k,j}$ from~(\ref{eq:deltaopt}), the small-variation algorithm calculates $\Delta^{Lin}_{k,j}$ defined as:
\begin{align}\label{eq:deltalin}
\Delta^\text{Lin}_{k,j}=\underset{\Delta_{k,j}}{\argmin}[||\bold{\epsilon}_{k,j-1}+\bold{F}_{k,j}^{Lin}||_2]
\end{align}
The objective function $||\bold{y}(\bold{x_\textit{k,j-1}})-\bold{s}+\bold{F_\textit{k,j}^\text{NL}}||_2$ in~(\ref{eq:perfect}) is approximated by a second order equation, given by $||\bold{\epsilon}_{k,j-1}+\bold{F}_{k,j}^{Lin}||_2$. \color{black}Using partial derivatives, the optimum value of $\Delta_{k,j}$ that minimizes~(\ref{eq:deltalin}), can be found by solving a system of two linear equations with two unknowns (the real and imaginary parts of $\Delta_{k,j}$), which makes the calculation much easier than minimizing the exact non-linear equation.
The main difficulty raised by the proposed algorithm is the complexity to assess the parameters $\bold{A}_{k,j}(m_1,m_2)$ as they depend on all Volterra coefficients. 
Section IV will be devoted to this question.

\subsection{Linearity Assumption}\label{sec:LA}
The variation $\Delta^\text{Lin}_{k,j}$ is computed based on the approximation~(\ref{eq:approx sv}), which is only valid for small values of  $\Delta^\text{Lin}_{k,j}$. \color{black}
In practice, we consider that the applied variation $\Delta^\text{applied}_{k,j}$ has at least to decrease the Euclidian distance between the initial and the received symbols. 
Mathematically, this is expressed as:
\begin{align}\label{eq:condit}
(||\bold{\epsilon}_{k,j-1}+\bold{F}_{k,j}^\text{NL}||_2\big{|}\Delta_{k,j}=\Delta^\text{applied\color{black}}_{k,j})\leq||\bold{\epsilon}_{k,j-1}||_2.
\end{align}
Taking $\Delta^\text{applied}_{k,j}=\Delta^\text{Lin}_{k,j}$ does not ensure that~(\ref{eq:condit}) is verified at each step since the linear assumption may not be met. Therefore, we consider instead that the applied variation is given by:
\begin{align}
\Delta^\text{applied}_{k,j}=\gamma \Delta^\text{lin}_{k,j},
\end{align}
where $\gamma$ is a real number in the interval $[0,1]$. It is proven in Appendix B that $\gamma \Delta^\text{lin}_{k,j}$ is a sub-optimum solution of the second order approximation of the objective function:
\begin{align}\label{annexb}
(||\bold{\epsilon}_{k,j-1}+\bold{F}_{k,j}^\text{lin}||_2\big{|}\Delta_{k,j}=\gamma\Delta^\text{Lin}_{k,j})\leq||\bold{\epsilon}_{k,j-1}||_2.
\end{align}
It is always possible to define $\gamma$ small enough to meet the linear approximation~(\ref{eq:approx sv}), so that the sub-optimum solution of~(\ref{eq:deltalin}) becomes also a sub-optimum solution of~(\ref{eq:perfect}), which means that $\Delta^\text{applied}_{k,j}$ satisfies~(\ref{eq:condit}). 
The value of $\gamma$ could be optimized at each step of the algorithm.
For instance, decreasing values of $\gamma$ can be applied until~(\ref{eq:condit}) is true. 
However, the complexity of such approach is difficult to predict.
In this work, we follow an approach similar to the trust-region method described in~\cite{num_opt}, where the norm of the applied variation $|\Delta^\text{applied}_{k,j}|$ is limited to a pre-defined value $\Delta_{max}$. 
The value of $\gamma$ is chosen so as to make this statement true. 
Mathematically, $\gamma$ is defined as follows:
\begin{align}
\gamma=\begin{cases}&1,\hspace{5mm}|\Delta^\text{lin}_{k,j}|\le\Delta_{max},\\&\Delta_{max}|\Delta^\text{lin}_{k,j}|^{-1},\hspace{5mm}|\Delta^\text{lin}_{k,j}|>\Delta_{max}.\end{cases}
\end{align}
If the so obtained $\gamma$ and the resulting $\Delta^\text{applied}_{k,j}$ does not meet~(\ref{eq:condit}), no variation is applied at the given step. The value of $\Delta_{max}$ is a trade-off between convergence speed and maximum achievable performance, as shown in Section~\ref{sec:results}. 
\subsection{Linear filtering}\label{sec:linear filtering}
Besides the iterative pre-distortion algorithm, a linear filter is applied to the transmitted signal to remove the linear interference caused by the channel.
The iterative pre-distortion algorithm includes this linear filter as part of the channel.
Intuitively this improves the convergence of the pre-distortion algorithm because the optimum values of the pre-distorted symbols are closer to the values of the un-pre-distorted symbols.
Note that this filter could alternatively be applied at the receiver.
This would keep the peak-to-average power ratio (PAPR) lower at the transmitter and limit the induced higher-order terms (see~\cite{principles}). 
However, this would also amplify the noise on the channel deeps.
In this work, we stick to the transmitter alternative.
 \color{black}
\section{Complexity Reduction of the Small-Variation Algorithm}
\label{sec:complex}
The algorithm presented in the previous section is rather theoretical, as the number of Volterra coefficients can be very large. We now present more practical methods to calculate the coefficients $A^{n}_{k,j}(m_1,m_2)$. The coefficients $A^{n}_{k,j}(1,0)$ and $A^{n}_{k,j}(0,1)$ are necessary to calculate $\Delta^\text{Lin}_{k,j}$. We propose three methods to compute these coefficients. The first has arbitrarily high precision, so that it can be considered as an exact and practical method to calculate $A^{n}_{k,j}(1,0)$ and $A^{n}_{k,j}(0,1)$. The two other methods calculate only approximations of these coefficients, but are less complex than the first method. The other coefficients $A^{n}_{k,j}(m_1,m_2)$, with $m_1>1$ or $m_2>1$, have also to be computed in order to verify the linearity assumption, so that the complexity still remains very high. The final part of this section shows how the calculations of these coefficients can be avoided.

\subsection{Calculations of the Coefficients $A^{n}_{k,j}(1,0)$ and $A^{n}_{k,j}(0,1)$ Based on Channel Simulations}
The coefficients $A^{n}_{k,j}(1,0)$ and $A^{n}_{k,j}(0,1)$ can be estimated  by the following procedure. At Step $j$ of Iteration $k$, the channel outputs are calculated that result from three different channel input variations:
\begin{align}\label{eq:variations}
1) \Delta_{k,j}=0\hspace{15mm} 2) \Delta_{k,j}=\epsilon_\text{r} \hspace{15mm} 3) \Delta_{k,j}=\epsilon_\text{i}
\end{align}
where $\epsilon_\text{r}$ and $\epsilon_\text{i}$ are respectively small real and pure imaginary numbers. The respective channel outputs are denoted as $y(n)$, $y_\text{r}(n)$ and $y_\text{i}(n)$.
If $\epsilon_\text{r}$ and $\epsilon_\text{i}$ are chosen sufficiently small, there is a linear relation between the input and output variations:
\begin{align} \label{eq:Rel}
&y_\text{r}(n)=y(n)+A^{n}_{k,j}(1,0)\epsilon_\text{r}+A^{n}_{k,j}(0,1)\epsilon_\text{r}\nonumber\\
&y_\text{i}(n)=y(n)+A^{n}_{k,j}(1,0)\epsilon_\text{i}-A^{n}_{k,j}(0,1)\epsilon_\text{i}.
\end{align}
For each $n$,~(\ref{eq:Rel}) form a set of two equations with two unknowns, $A^{n}_{k,j}(1,0)$ and $A^{n}_{k,j}(0,1)$, so that they can easily be estimated. Only $L_c$ sets of equations need to be solved due to the finite channel length assumption. The smaller $\epsilon_\text{r}$ and $\epsilon_\text{i}$, the more accurate the calculation. This method still has a high complexity since three simulations of the channel at each step of each iteration. \\%
To simulate the channel, the pre-distortion block needs to oversample the signal by several times the symbol rate in order to avoid spectral aliasing from the non-linear interference. 
For channels with large memory, the channel simulations can require a high complexity. In the following sections, we describe methods to obtain approximations of the coefficients $A^{n}_{k,j}(1,0)$ and $A^{n}_{k,j}(0,1)$ with a lower complexity.

\subsection{Calculations of the Coefficients $A^{n}_{k,j}(1,0)$ and $A^{n}_{k,j}(0,1)$ Based on a Reduced Volterra Model}
\label{sec:volt}
The coefficients $A^{n}_{k,j}(1,0)$ and $A^{n}_{k,j}(0,1)$ can be computed by summing the contribution of each Volterra coefficient. 
To decrease the algorithm complexity, they can instead be approximated by summing the contributions of only the most significant Volterra coefficients. 
Let us consider the generic Volterra coefficient $H_{2m+1}(n_1...n_{2m+1})$. Volterra coefficients that do not have at least one index equal to $n-j$ can be neglected, since only the symbol $j$ is modified. 
To further decrease the number of Volterra coefficients, an approximated value of $A^{n}_{k,j}(m_1,m_2)$ can be calculated by truncating the non-linearity order and by limiting the channel length. 
Truncating the channel length to a given value $L^{'}_\text{c}$ implies that only the Volterra coefficients $H_{2m+1}(n_1...n_{2m+1})$ for which each index $n_i$ satisfies $|n_i|\leq L^{'}_c$ are considered. 
Moreover, $A^{n}_{k,j}(m_1,m_2)=0$ for $|n-j|>L^{'}_\text{c}$. 
Different approximations are proposed in~\cite{polynomials} to further decrease the number of coefficients. In this paper, we further reduce the number of Volterra coefficients by only considering the ones depending on maximum $2$ different indexes.
\subsection{Calculations of the Coefficients $A^{n}_{k,j}(1,0)$ and $A^{n}_{k,j}(0,1)$ Using a Look-Up Table} 
\label{sec:lookup}
The idea of this approximation is to pre-compute the values of $A^{n}_{k,j}(1,0)$ and $A^{n}_{k,j}(0,1)$ and to store them in a look-up table. Since $A^{n}_{k,j}(1,0)$ depends on the symbols $[x_{k,j-1}(n-L_2)$ ... $x_{k,j-1}(n+L_1)]$, which take on continuous values, an infinite number of pre-computed table entries would need to be stored. Therefore, an approximation of $A^{n}_{k,j}(1,0)$ is calculated by rounding each value in $[x_{k,j-1}(n-L_2)$ ... $x_{k,j-1}(n+L_1)]$ to the closest value in $C$, where $C=\{c_1, c_2...c_P\}$ is a set of $P$ complex numbers. Considering the channel length $L^{'}_\text{c}$, approximately $P^{L^{'}_\text{c}}$ values need to be stored. To avoid the complexity of rounding each pre-distorted symbol to the closest value in $C$, a further approximation can be introduced, considering that the coefficients $A^{n}_{k,j}(1,0)$ and $A^{n}_{k,j}(0,1)$ are independent of $k$. This means that these coefficients are calculated at Iteration $k$ using the symbols $[s(n-L_2)$ ... $s(n+L_1)]$ instead of the symbols $[x_{k,j-1}(n-L_2)$ ... $x_{k,j-1}(n+L_1)]$.

\subsection{Alternative to the Calculation of the Remaining Coefficients $A^{n}_{k,j}(m_1,m_2)$}
\label{sec:alter}
Once the precoded symbols are updated using the lower complexity algorithms proposed in Sections \ref{sec:volt}~and~\ref{sec:lookup}, it is important to check the linearity assumption based on which the algorithms rely. This can be simply done by simulating the actual channel, but we wanted to avoid that in Sections \ref{sec:volt}~and~\ref{sec:lookup} for complexity reasons.
 A pragmatic approach to keeping the low-complexity advantage is to check the linearity assumption only at the end of each iteration by a single channel simulation. We consider that the linearity assumption is met if the MSE decreases after each iteration of the algorithm. When the Euclidian distance stops decreasing, the algorithm is stopped and the pre-distorted values of the previous iteration are kept as the final values.

\subsection{Complexity Comparison}\label{sec:compl}
In this subsection, the complexity of the small-variation algorithm is compared to that of its reduced-complexity alternatives. In the following, $\Delta_{k,j}$ and $\Delta^*_{k,j}$ are replaced by $\Re(\Delta_{k,j})$ and $\Im(\Delta_{k,j})$ in ($\ref{eq:approx sv}$), and the coefficients $A^{n}_{k,j}(1,0)$ and $A^{n}_{k,j}(0,1)$ are accordingly modified. This simple notational change allows a (small) complexity decrease. The algorithm consists in $K$ iterations of $N$ steps. At each step, the initial algorithm and the reduced-complexity algorithms all require the following operations:
\begin{itemize}
\item{}
Calculate the coefficients $A^{n}_{k,j}(1,0)$ and $A^{n}_{k,j}(0,1)$ from~($\ref{eq:approx sv}$). These coefficients are calculated differently for each method.
\item{}
Evaluate~$||\bold{\epsilon}_{k,j-1}+\bold{F}_{k,j}^{Lin}||_2$ in~($\ref{eq:deltalin}$)\color{black}, with $\bold{F}_{k,j}^\text{NL}=\bold{F}_{k,j}^\text{Lin}$ \color{black} as defined in~($\ref{eq:yoyo}$). The computation of the norm of this vector is obtained by summing the norm of each element of the vector $\bold{\epsilon}_{k,j-1}+\bold{F}_{k,j}^{Lin}$. \color{black}It can be shown that this necessitates $10L_\text{c}$ multiplications and $5(L_\text{c}-1)$ additions. 
\item{}
Find the minimum of~($\ref{eq:deltalin}$). Using partial derivatives, this requires the inversion of a $2\times 2$ (real) matrix, $4$ multiplications, and $2$ additions.
\item{}
Update the channel outputs for the next step (in fact calculate $\bold{\epsilon_{k,j}}$ from $\bold{\epsilon_{k,j-1}}$ using $A^{n}_{k,j}(1,0)$, $A^{n}_{k,j}(0,1)$ and the applied variation). This requires $4L_c$ multiplications and $4L_c$ additions.
\end{itemize}
The complexity to determine the linear coefficients depends on the chosen algorithm:
\begin{itemize}
\item{}
The complexity of the algorithm using the Volterra model depends on the number of considered Volterra coefficients and on the order of the Volterra coefficient. A Volterra coefficient of order $p$ requires $4p$ multiplications and 2 additions. By truncating the channel length to $L^{'}_\text{c}$, each of the $L^{'}_\text{c}$ outputs depends on approximately $(L^{'}_\text{c})^{(p-1)}$ coefficients of order $p$.
\item{}
The method based on channel simulations necessitates three channel simulations. At each channel simulation, $L_\text{c}$ outputs need to be calculated. This requires simulating the power amplifier output for more than $L_c k_\text{OSF}$ input values, where $k_\text{OSF}$ is the oversampling factor. The output of the power amplifier can be obtained by interpolating the AM-AM and AM-PM characteristics or assessed using a polynomial approximation. The number of operations to calculate the power amplifier output is then approximatively proportional to the chosen non-linearity order. The complexity of the convolution with the linear filters also needs to be taken into account, which is proportional to $L_\text{c}^2 k_\text{OSF}$. To calculate the coefficients $A^{n}_{k,j}(1,0)$ and $A^{n}_{k,j}(0,1)$ from the channel simulations, $4$ additions and $4$ divisions need to be done.
\item{}
The method based on look-up tables needs to round $4L^{'}_\text{c}$ values to the closest value in the look-up table since the $L^{'}_\text{c}$ complex outputs depend on $2L^{'}_\text{c}$ inputs.
\end{itemize}
The method based on the look-up tables is the least complex. The method based on the approximated Volterra model is less complex than the method based on channel simulations only if the number of Volterra coefficients is very small, so that very short channel lengths must be considered. This will be further discussed in the next section.

\section{Numerical results}\label{sec:results}
We consider $32$-APSK symbols and SRRC shaping and receiver filters. 
If not specified differently, the roll-off factor is assumed to be equal to $0.1$.  
\begin{figure}
\center
\includegraphics[width=\columnwidth]{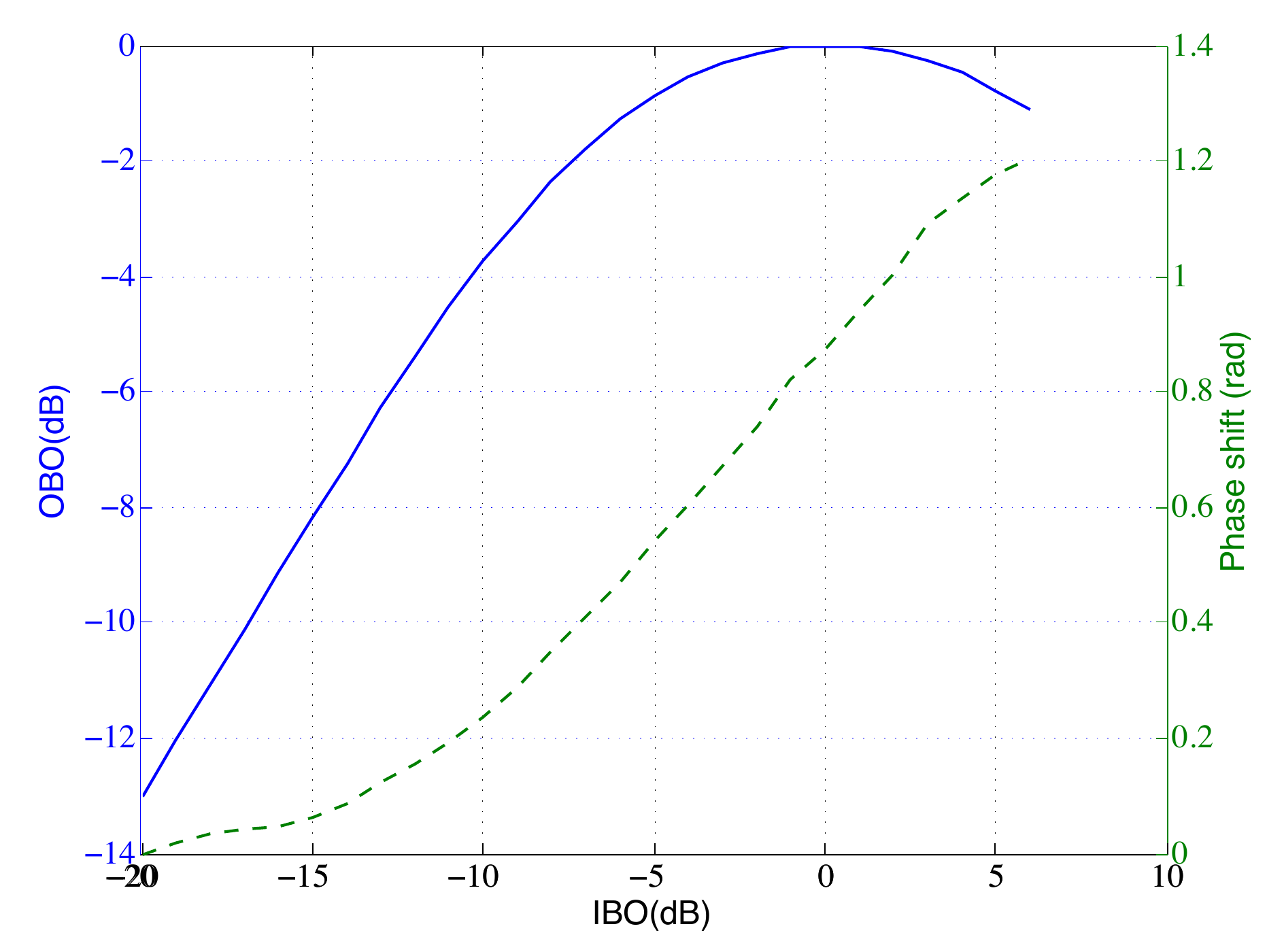}
\caption{AM-AM and AM-PM characteristics of the HPA.}
\label{fig2}
\end{figure}
\begin{figure}
\center
\includegraphics[width=0.7\columnwidth]{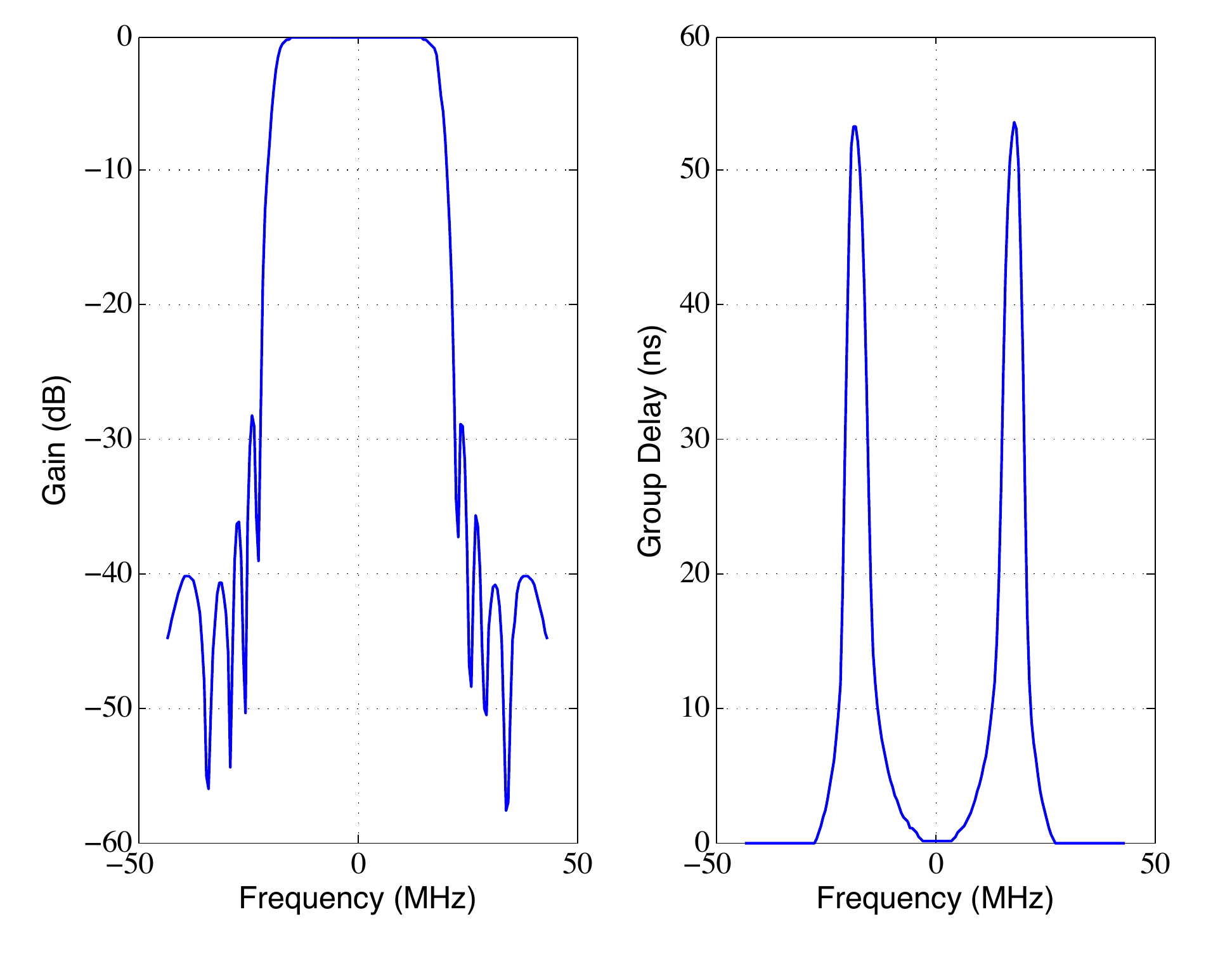}
\caption{IMUX characteristics.}
\label{fig3}
\end{figure}
\begin{figure}
\center
\includegraphics[width=0.7\columnwidth]{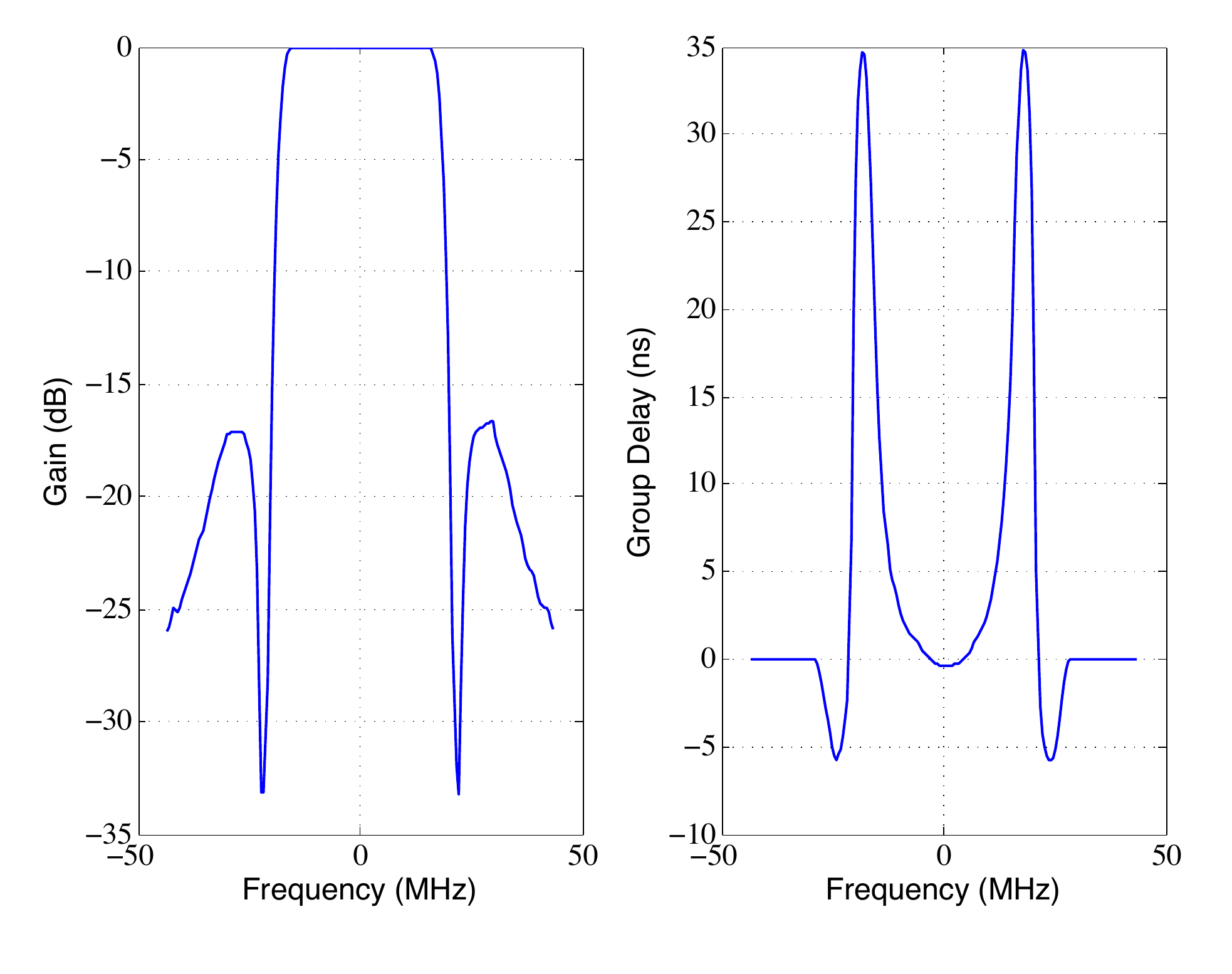}
\caption{OMUX characteristics.}
\label{fig3bis}
\end{figure}
A traveling-wave tube (TWT) amplifier is considered with AM-AM and AM-PM characteristics given in Fig.~\ref{fig2}. The IMUX and OMUX characteristics are given in Fig.~\ref{fig3} and Fig.~\ref{fig3bis} respectively. 
Their $3$-dB cut-off frequency is equal to $36$ MHz. 
The low density parity-check (LDPC) encoder and the interleaver are the ones defined in the DVB-S2 standard for the 32-APSK modulation and the code rate equal to 3/4. 
As discussed in Section~\ref{sec:MED}, $N$ is equal to the number of symbols in a PLFRAME of the DVB-S2 standard, which is equal to $12960$ for the case of 32-APSK modulation. 
The symbol rate is equal to $36$MBauds, so that the channel occupation is equal to $110\%$ of the theoretical channel bandwidth. This allows increasing the spectral efficiency at the cost of more non-linear interference.
\begin{figure}
\center
\includegraphics[width=\columnwidth]{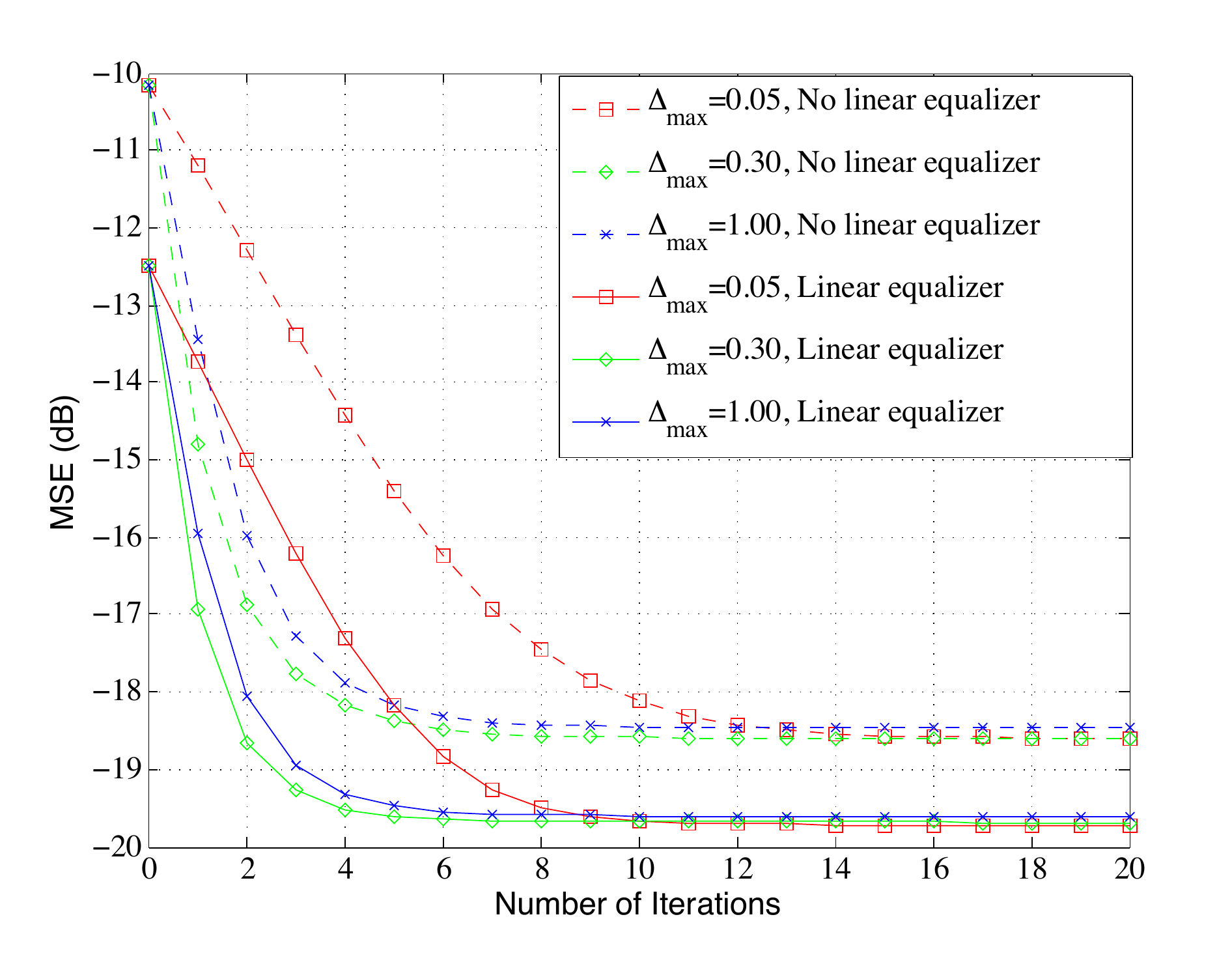}
\caption{Mean-square error (MSE) after each iteration of the small-variation algorithm, symbol rate$=36$ MSymb/s.}
\label{fig4}
\end{figure}
Fig.~\ref{fig4} illustrates the MSE between the initial and the received symbols after each iteration of the small-variation algorithm described in Section~\ref{Sec3} for different values of $\Delta_{max}$, with or without a linear zero-forcing filter. 
Fig.~\ref{fig4} shows that a better optimum is reached when a linear zero-forcing filter is used so that it will always be considered in the following results. 
The linear zero-forcing filter is placed at the transmitter. 
At each step of the algorithm, it is checked that the variation $\Delta^\text{applied}_{k,j}$ decreases the square error so that the algorithm always converges to a local optimum.
The value $\Delta_{max}$ has no impact on the value of the local optimum but controls the convergence speed of the algorithm. 
Too small values of $\Delta_{max}$ obviously decrease the speed of convergence of the algorithm. 
This is also the case for too large values of $\Delta_{max}$ since it increases the number of steps where no variation is applied.
\begin{figure}
\center
\includegraphics[width=\columnwidth]{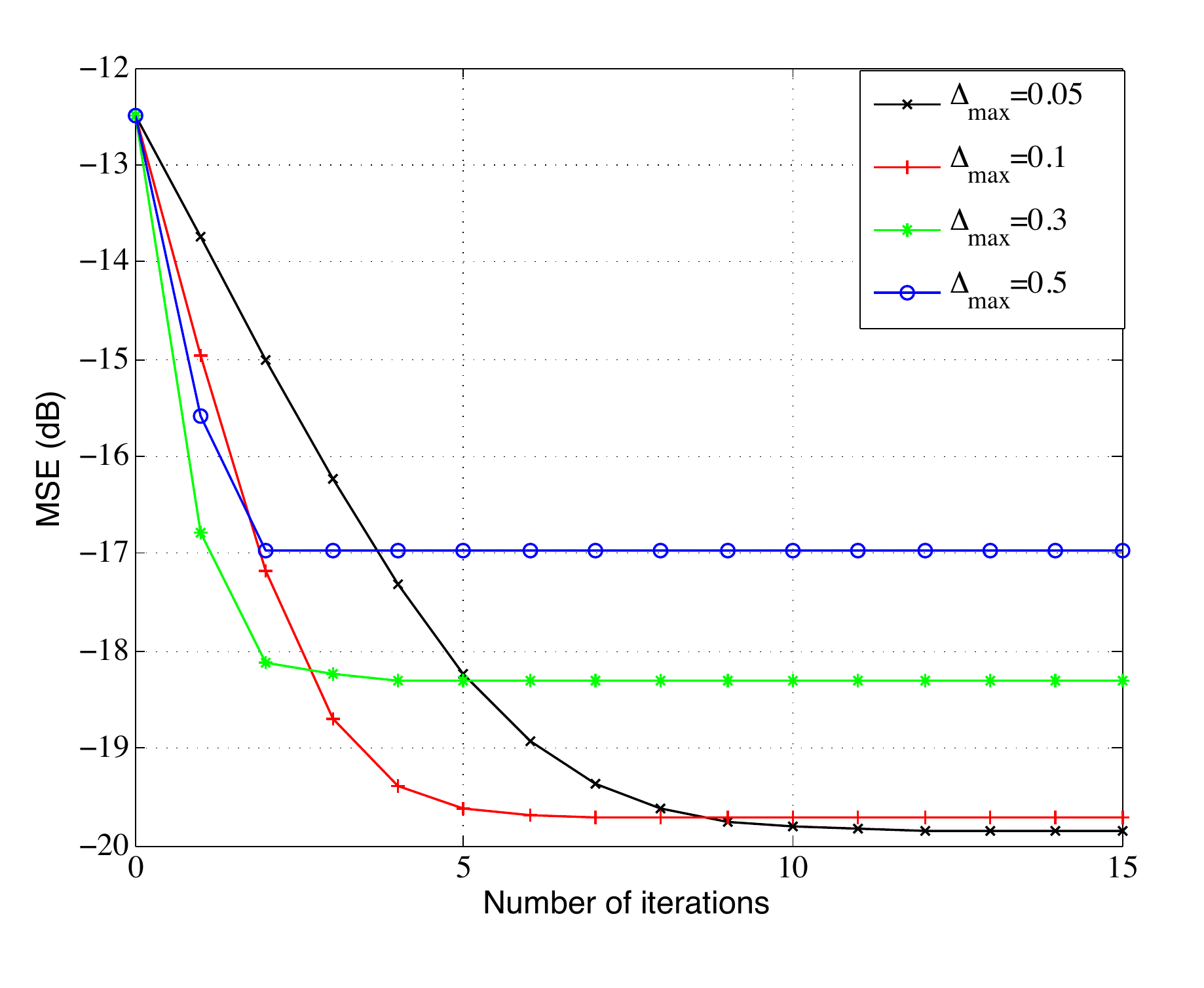}
\caption{Mean-square error (MSE) after each iteration of the small-variation algorithm, using the convergence method described in Section~\ref{sec:alter}, symbol rate$=36$ MSymb/s.}
\label{fig5}
\end{figure}
Fig.~\ref{fig5} illustrates the performance reached when the square error decrease is only checked at the end of each iteration (instead of the end of each step) as described in Section~\ref{sec:alter}. 
It can be observed that the asymptotic performance now depends on the considered $\Delta_{max}$. 
Sufficiently small values of $\Delta_{max}$ ($0.05$ and $0.1$) however allow to reach similar performance as in previous figure. 
\begin{table*}[ht]
\renewcommand{\arraystretch}{1.3}
\centering
\caption{Performance of the reduced-complexity alternatives of the small-variation algorithm.}
\label{table2}
\begin{tabular}{|c|c|c|c|}
\hline
&IBO=$3$dB & IBO=$4$dB & IBO=$5$dB  \\
\hline
MSE small-variation algorithm (reference)& $\text{Ref}=-19.71dB$&$\text{Ref}=-20.72dB$&$\text{Ref}=-22.22dB$\\
\hline
Look-up table, $L^{'}_c=3$&$=\text{Ref}+0.78dB$&$=\text{Ref}+0.36dB$&$=\text{Ref}+0.18dB$\\
\hline
Look-up table, $L^{'}_c=5$&$=\text{Ref}+0.43dB$&$=\text{Ref}+0.08dB$&$=\text{Ref}+0.02dB$\\
\hline
Reduced-Volterra model, $L^{'}_c=3$& $=\text{Ref}+1.10dB$&$=\text{Ref}+0.76dB$&$=\text{Ref}+0.65dB$\\
\hline
Reduced-Volterra model, $L^{'}_c=5$& $=\text{Ref}+0.78dB$&$=\text{Ref}+0.45dB$&$=\text{Ref}+0.35dB$\\
\hline
\end{tabular}
\end{table*}

Table~\ref{table2} shows the performance loss associated to the reduced-complexity alternatives of the small-variation algorithm, proposed in Section~\ref{sec:volt} and~\ref{sec:lookup}. 
For both approximations, the performance loss decreases with the IBO. Considering $L^{'}_c=5$ allows a small performance increase compared to $L^{'}_c=3$. 
Slightly better performance is achieved with the method based on look-up tables, which necessitates a look-up table of $L^{'}_c\times32^{L^{'}_c}$ entries. \color{black}
The method based on a reduced Volterra model however does not require any pre-computation and still allows a decrease in complexity compared to the method based on channel simulations. 
Considering $L^{'}_c=3$, each step of each iteration requires about $60$ multiplications and $30$ additions, relying on the fact that some products of the pre-distorted symbols can be reused to calculate different channel outputs. 
Using channel simulations with $L'_\text{c}=3$ and an oversampling factor equal to 8, $40$ samples need to be filtered by the shaping and IMUX filters, interpolated, and re-filtered by the OMUX filter and the receiver filter. 
Considering that the shaping and IMUX filters (and OMUX and receiver filters) have an impulse response longer than $40$ samples, this means that already $40^2$ multiplications are involved for each convolution. 
Clearly, the complexity is lower using the Volterra model instead of channel simulations.
\begin{figure}
\center
\includegraphics[width=\columnwidth]{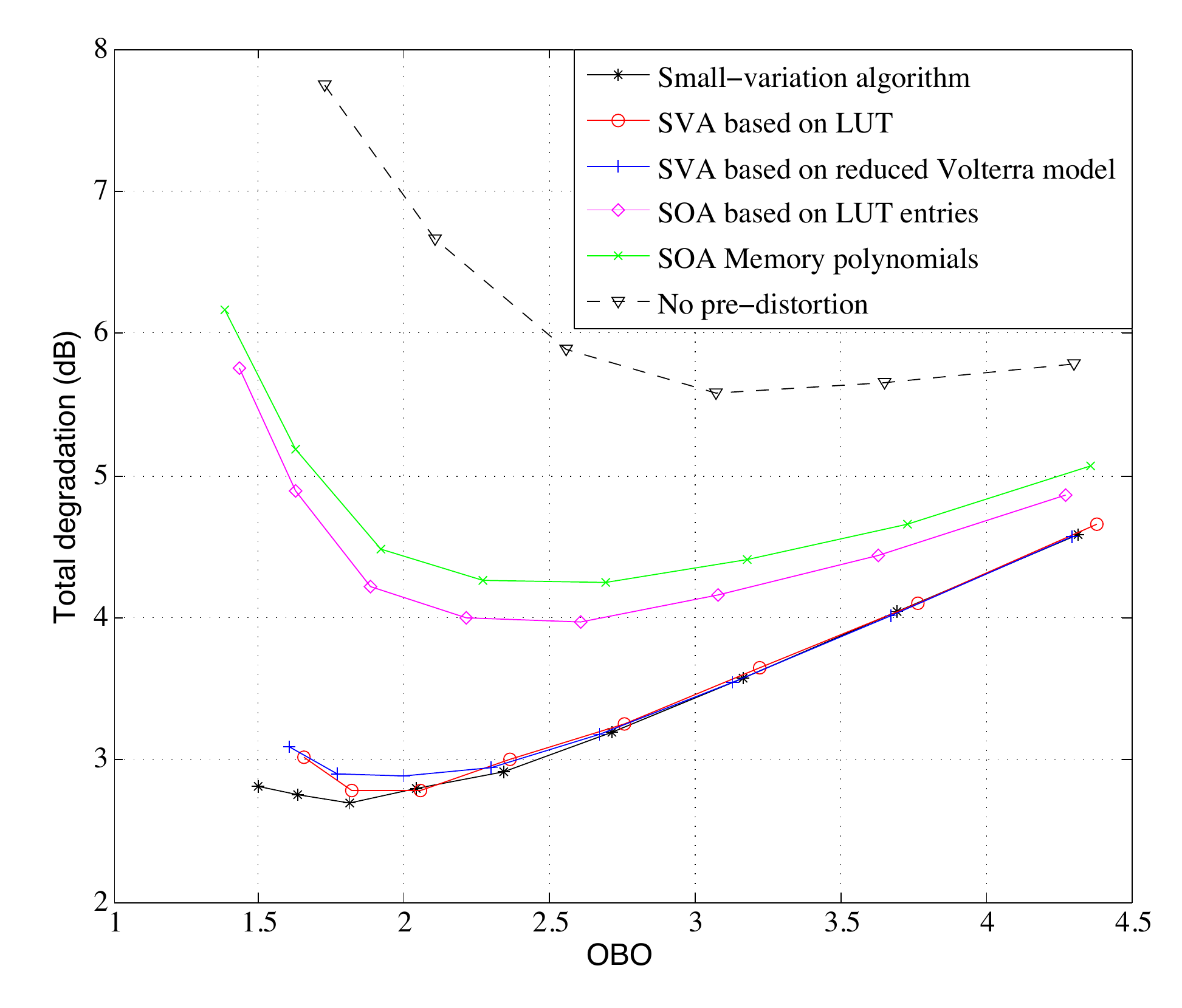}
\caption{Total degradation for small-variation algorithm (SVA) and state-of-the-art pre-distortion methods, symbol rate$=36$ MSymb/s, roll-off$=0.1$.}
\label{fig7}
\end{figure}
\\Fig.~\ref{fig7} compares the small-variation algorithm and its reduced-complexity alternatives to state-of-the-art algorithms. 
The comparison is performed based on the total degradation, described in Section~\ref{sec:td}, as a function of the OBO. The target BER is equal to $10^{-5}$. 
Two state-of-the-art methods are considered for the comparison. 
The first method is based on memory polynomials, where the pre-distorter is a reduced Volterra system presented in~\cite{polynomials}. 
Third order Volterra coefficients and a pre-distorter length $L^{'}_c=9$ have been considered. 
The second method is the one proposed in~\cite{Casini:modem}, where the value of each pre-distorted symbol is a function of the neighboring un-pre-distorted symbols. 
All possible combinations are pre-computed offline and stored in a look-up table. 
A pre-distorter length $L^{'}_c=3$ is considered so that $32^3$ entries are stored in the look-up table. Among all pre-distortion methods, this approach has the lowest real-time complexity, since only one memory access per symbol needs to be performed. Moreover, Fig.~\ref{fig7} shows that it outperforms the first state-of-the art method in the considered scenario. 
Fig.~\ref{fig7} also shows that the small-variation algorithm and the reduced-complexity alternatives outperform the state-of-the-art algorithms. About $1.2$dB is gained on the optimum total degradation point. 
The performance of the reduced-complexity alternatives of the small-variation algorithm is assessed assuming also a pre-distorter length equal to $L^{'}_c=3$. 
The loss of the reduced-complexity alternatives on the optimum total degradation point is on the other hand smaller than $0.2$dB.\\
\begin{figure}
\center
\includegraphics[width=\columnwidth]{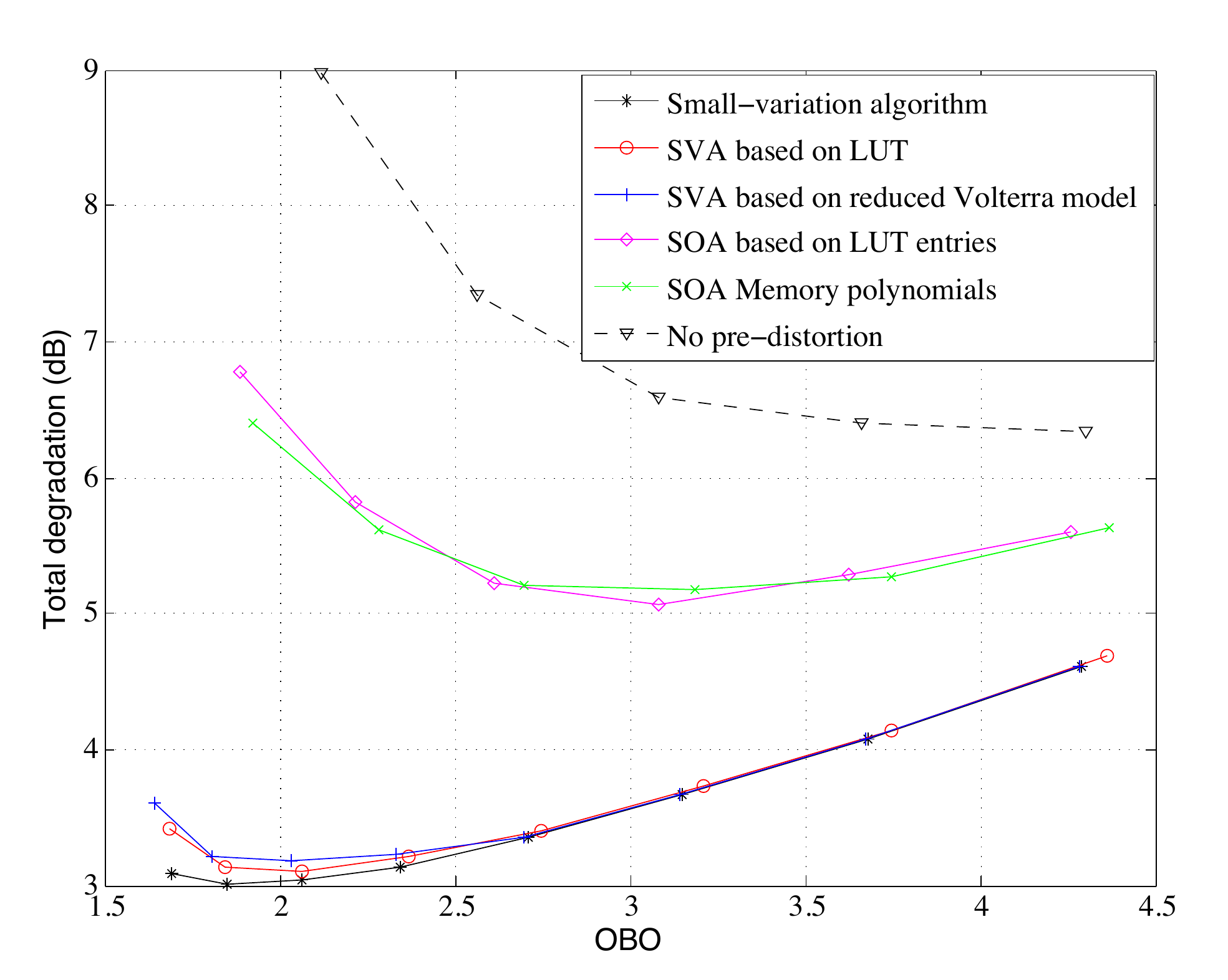}
\caption{Total degradation for small-variation algorithm (SVA) and state-of-the-art pre-distortion methods, symbol rate$=38$ MSymb/s, roll-off$=0.05$.}
\label{fig8}
\end{figure}

Fig.~\ref{fig8} considers an increased symbol rate equal to $38$MHz and a reduced roll-off factor equal to $0.05$, simulating therefore a higher interference scenario. 
As a result, the total degradation using the small-variation algorithm is higher for every OBO when compared to the previous case. The 2 state-of-the-art methods give similar performance. 
The gain compared to these state-of-the-art algorithms is higher than in previous case and is about $2$dB. 
The reduced-complexity alternatives reach again almost the same optimum total degradation.
\color{black}

 \section{Conclusion and future work}
This paper proposes a new iterative pre-distortion algorithm suited to the use of high-order modulations on a highly non-linear satellite communication channel. 
The algorithms aims at minimizing the Euclidian distance between the transmitted and received symbols. 
The pre-distorted symbols are updated at each iteration based on a linear approximation of the channel output variation, which is only valid if the symbol update is kept sufficiently local.
However, a major issue of the algorithm is the complexity involved in the estimation of the linear relation between the channel input and output variations. 
Two approximations have been proposed to strongly decrease the algorithm complexity. 
The first one relies on a reduced Volterra model and the second one is based on the use of look-up tables. 

The performance improvement brought by the algorithm, compared to state-of-the-art algorithms, represents several dBs on the MSE and $1$ to $2$dBs on the link budget with $32$-APSK modulation. The channel occupancy bandwidth is equal to $110\%$ of the theoretical channel bandwidth to improve the spectral efficiency. Roll-off factors as low as $0.05$ have been considered. The performance improvement is obtained at the cost of a complexity increase. 
The choice of the pre-distortion algorithm is therefore a performance/complexity trade-off. \color{black}
Future work will include the algorithm extension to the case where more than one carrier is amplified by the same power amplifier. 
If the channel is known and if all signals are transmitted from the same hub, \color{black}a pre-distortion algorithm similar to the one proposed here for a single carrier per channel can be applied.
\begin{appendices} 
\section{Calculation of the coefficients $A^{n}_{k,j}(m_1,m_2)$ for some simple Volterra models}
Let us first consider a channel consisting only of the third-order Volterra coefficient $H_3(0,0,0)$.  Each element of the difference between the output $\bold{y}(\bold{x_{k,j-1}}+\bold{\Delta_{k,j}})$ and the output $\bold{y}(\bold{x_{k,j-1}})$ is given by:
\begin{align} \label{eq:ex 3}
F^\text{NL}_{k,j}(j)&=H_3(0,0,0)\{|x_{k,j-1}(j)+\Delta_{k,j}|^2[x_{k,j-1}(j)+\Delta_{k,j}]\nonumber\\&\hspace{23mm}-|x_{k,j-1}(j)|^2x_{k,j-1}(j)\}\nonumber\\
&=H_3(0,0,0)[x_{k,j-1}(j)^2\Delta_{k,j}^*+2|x_{k,j-1}(j)|^2\Delta_{k,j}\nonumber\\
&\hspace{-2mm}+2x_{k,j-1}(j)|\Delta_{k,j}|^2+x_{k,j-1}(j)^*\Delta_{k,j}^2+|\Delta_{k,j}|^2\Delta_{k,j}].
\end{align}
The other outputs are not modified since the channel is memoryless. The different coefficients $A^{n}_{k,j}(m_1,m_2)$ can be directly estimated from~(\ref{eq:ex 3}), and are given in the second column of Table~\ref{table co}. 
\begin{table*}[ht]
\renewcommand{\arraystretch}{1.3}
\centering
\caption{Coefficients $A^{n}_{k,j}(m_1,m_2)$ for the Fictive Model $H_3(0,0,0)$, $H_3(1,0,0)$ and $H_3(0,1,2)$.}
\label{table co}
\begin{tabular}{|c|c|c|c|}
\hline
Coefficient & Value for $H_3(0,0,0)$ & Value for $H_3(1,0,0)$ & Value for $H_3(0,1,2)$ \\
\hline
$A_{k,j}^j(1,0)$& $2H_3(0,0,0)|x_{k,j-1}(j)|^2$&$H_3(1,0,0)x_{k,j-1}(j-1)x_{k,j-1}(j)^*$&$H_3(0,1,2)x_{k,j-1}(j-1)x_{k,j-1}(j-2)^*$\\
$A_{k,j}^j(0,1)$& $H_3(0,0,0)x_{k,j-1}(j)^2$&$H_3(1,0,0)x_{k,j-1}(j-1)x_{k,j-1}(j)$&$0$\\
$A_{k,j}^j(2,0)$&$H_3(0,0,0)x_{k,j-1}(j)^*$&$0$&$0$\\
$A_{k,j}^j(1,1)$& $2H_3(0,0,0)x_{k,j-1}(j)$&$H_3(1,0,0)$&$0$\\
$A_{k,j}^j(2,1)$& $H_3(0,0,0)$&$0$&$0$\\
\hline
$A^{k,j+1}_j(1,0)$& $0$&$H_3(1,0,0)|x_{k,j-1}(j+1)|^2$&$H_3(0,1,2)x_{k,j-1}(j-1)x_{k,j+1}(j-1)^*$\\
\hline
$A^{k,j+2}_j(0,1)$& $0$&0&$H_3(0,1,2)x_{k,j+1}(j-1)x_{k,j+2}(j-1)$\\
\hline
others&$0$&$0$&$0$\\
\hline
\end{tabular}
\end{table*}
The third and fourth columns of Table~\ref{table co} give the non-zero values for $A^{n}_{k,j}(m_1,m_2)$, considering a channel with a single Volterra coefficient respectively equal to $H_3(1,0,0)$ and $H_3(0,1,2)$. It should be noticed that more than one input is now modified, due to the memory of the system. Moreover, the output variation for $H_3(0,1,2)$ is linear in $\Delta_{k,j}$ because all indexes of this Volterra coefficient are different.
\section{Proof of~(\ref{annexb})}
By definition of $\Delta^\text{Lin}_{k,j}$, we have that:
\begin{align}\label{eq:ega}
&[||\bold{\epsilon}_{k,j-1}+\bold{F}_{k,j}^\text{Lin}||_2\big{|}\Delta_{k,j}=\Delta^\text{Lin}_{k,j}]=\nonumber\\&||\bold{\epsilon}_{k,j-1}||_2+\sum_n|A^n_{k,j}(1,0)\Delta^\text{Lin}_{k,j}+A^n_{k,j}(0,1)(\Delta^\text{Lin}_{k,j})^*|^2\nonumber\\
&\hspace{5mm}+2\Re{\{\epsilon_{k,j-1}(n)^*[A^n_{k,j}(1,0)\Delta^{Lin}_{k,j}+A^n_{k,j}(0,1)(\Delta^{Lin}_{k,j})^*]\}}\nonumber\\&\leq ||\bold{\epsilon}_{k,j-1}||_2.
\end{align} 
Therefore, we have that the second line of~(\ref{eq:ega}) is negative and that:
\begin{align}\label{eq:logique}
&\sum_n|A^n_{k,j}(1,0)\Delta^{Lin}_{k,j}+A^n_{k,j}(0,1)(\Delta^{Lin}_{k,j})^*|^2\nonumber\\&\leq-2\Re{[\epsilon^{Lin}_{k,j-1}(n))^*(A^n_{k,j}(1,0)\Delta_{k,j}+A^n_{k,j}(0,1)(\Delta^{Lin}_{k,j})^*]}.
\end{align} 
Since $\gamma^2<\gamma$, it is easy to see that:
\begin{align}\label{eq:ega2}
&[||\bold{\epsilon}_{k,j-1}+\bold{F}_{k,j}^\text{Lin}||_2\big{|}\Delta_{k,j}=\Delta^{\gamma}_{k,j}]=||\bold{\epsilon}_{k,j-1}||_2\nonumber\\&\hspace{5mm}+\gamma^2\sum_n|A^n_{k,j}(1,0)\Delta^\text{Lin}_{k,j}+A^n_{k,j}(0,1)(\Delta^\text{Lin}_{k,j})^*|^2\nonumber\\
&\hspace{5mm}+2\gamma\Re{\{\epsilon_{k,j-1}(n)^*[A^n_{k,j}(1,0)\Delta^{Lin}_{k,j}+A^n_{k,j}(0,1)(\Delta^{Lin}_{k,j})^*]\}}\hspace{10mm}\nonumber\\&\leq||\bold{\epsilon_{k,j-1}}||_2.
\end{align}
\end{appendices} 
\bibliographystyle{IEEEtranTCOM}
\bibliography{bibli}
\end{document}